\documentstyle[11pt,paspconf]{article}

\begin{document}

\title{Optical Diffuse Light in Clusters of Galaxies}
\author{Rosendo V\'{\i}lchez--G\'omez}
\affil{Space Telescope Science Institute, 3700 San Martin Drive, Baltimore, MD 21218, USA}

\begin{abstract}
 
I present here a review of the observed characteristics of the optical diffuse light in clusters, 
the possible sources of this light and some of the theories that try to explain the existence of 
big envelopes around the brightest cluster galaxies.

\end{abstract}


\section{Introduction}

The first reference that we can find in the literature about the diffuse light in 
cluster of galaxies was given by Zwicky (1951): ``One of the most interesting
discoveries made in the course of this investigation [in the Coma cluster] is the
observation of an extended mass of luminous intergalactic matter of very low surface
brightness. The objects which constitute this matter must be considered as the faintest
individual members of the cluster. [We report] the discovery of luminous intergalactic
matter concentrated generally and differentially around the center of the cluster
and the brightest (most massive) galaxies, respectively''. This is a perfect characterization of 
the optical diffuse light in cluster: extended, low surface brightness and around the
center of the cluster.

Zwicky was trying to settle three of the problems of the extragalactic astronomy at that moment:
(1) this luminous intergalactic matter can account for the dark matter needed in Coma if this cluster were virialized; 
(2) the shape of the luminosity function (a Gaussian, according to Hubble) is monotonely increasing with
decreasing brightness; 
and (3) the galaxies extend notably far away from their centers\footnote{Baum (1955) claims that ``galaxies
blend into one another with no vacant intergalactic gaps in between''.}. 

The characteristics of this diffuse matter published by Zwicky (1951, 1957, 1959) were qualitative: it has
an extension of around 150 kpc, the color index is rather blue and produces a local absorption of light of the 
order of six tenth of a magnitude.

The first published attempt to obtain a value for the surface brightness of the faint intergalactic matter in Coma
corresponds to de Vaucouleurs (1960). He reported an upper limit of 
B $> 29.5$ mag\,arcsec$^{-2}$ at $\langle r \rangle \simeq0^{\circ}\llap{$.$}9$. With this value, de Vaucouleurs reasons 
out that ``a stellar population composed exclusively of extreme red dwarfs of mass M$< 0.1$ M$_{\odot}$ and absolute 
magnitudes $M(pg) > +15$ would, in principle, give an M/L ratio of the order measured in Coma. While such stars are
known to exist in the neighborhood of the sun, it seems very difficult to admit that they could populate intergalactic
space with the required density and to the exclusion of all other stars of slightly greater mass"\footnote{Actually, 
Boughn \& Uson (1997), studying three rich Abell clusters, where they don't detect any anomalous reddening in the 
intracluster medium, conclude that no more than 2h$^{-1}$\% of the dark matter can be in the form of low mass 
($\sim 0.1$ M$_{\odot}$) subdwarfs or old disk dwarfs.}. Thus, de Vaucouleurs 
concludes that the mass of the intergalactic
matter is not enough to account for the mass value estimated through the virial theorem.

The next step in the first studies of the diffuse light in clusters corresponds to Matthews, Morgan 
\& Schmidt (1964). During the analysis of radio sources, they found near the center of a number of Abell's rich clusters, 
supergiant D galaxies with diameters 3--4 times as great as the ordinary lenticulars in the same clusters. They gave the
prefix ``c'' to these very large D galaxies, ``in a manner similar to the notation for supergiant stars in stellar
spectroscopy''. The reason for this remark is that I believe that there is not a real difference between the detection
of intracluster light or the halo of a cD. Whether this diffuse light is called the cD envelope or diffuse intergalactic
light is a matter of semantics. In fact, Oemler (1973) in his study of Abell 2670 where he traced a diffuse envelope to 
almost 1 Mpc says: ``An important question is the relation between this diffuse component and the central elliptical 
galaxy, the combination of which seems to produce  the cD galaxy''. Nevertheless, there are clusters without a 
cD in the center where a diffuse light has been detected in its central part, as it is the case in Cl 1613+31
(V\'{\i}lchez--G\'omez, Pell\'o \& Sanahuja 1994a,b)

Before the CCD detectors were widely used, most of the observations and study of the diffuse light in clusters was carried
out in the Coma cluster: Abell (1965); Gunn (1969); de Vaucouleurs \& de Vaucouleurs (1970); Welch \& Sastry (1971, 1972); 
Gunn \& Melnick (1975); Mattila (1977); Melnick, White \& Hoessel (1977); Thuan \& Kormendy (1977). 
There are also some studies in Virgo: Holmberg (1958); 
Arp \& Bertola (1969); de Vaucouleurs (1969). Finally, there are also studies of the
faint envelopes of elliptical and cD galaxies: Arp \& Bertola (1971); Baum (1973); Kormendy \& Bahcall (1974); 
Oemler (1973, 1976). I will consider here, basically, the problems associated to the use of CCD's in the study of the diffuse
light as well as the results obtained with that kind of detectors.

\section{Problems and Errors}

If we consider that the intracluster light is expected to be extremely faint, about 25--26 mag arcsec$^{-2}$ in a red filter
(if it represents 10 to 25\% of the total light in the center of an intermediate redshift cluster), it is easy to understand
how hard can be to obtain a reliable detection and analysis of the diffuse light in a cluster. We have to be sure that
our detection is not the result of spurious effects, such as instrumental scattering or contamination due to bright stars or
faint galaxies. I will comment some of this error sources:

\subsubsection{Instrumental Scattering.}

The diffuse light due to the mirrors of the telescope is the first source of parasitic light. If the cleanliness of the 
telescope optics is not correct enough, some of the results that we could ascribe to the intracluster light would be masked
or spoiled.

\subsubsection{Flat Fielding.} 

Our images must be cleaned of any kind of residual ghost image structures as well as free of fringing. A good level of 
flattening should be lower than 0.5\%. 

\subsubsection{Contamination due to Bright Stars.} As we are trying to obtain accurate surface brightness profiles at 25 
mag arcsec$^{-2}$ and lower, it is necessary an accurated removal of the halos of stars and bright clusters members. An 
unaccurate subtraction can alter the result in more than 0.5 mag arcsec$^{-2}$. It is also essential to check for the
possibility of contamination due to halos of stars located outside but near the field. Some comments about the 
removal of the halos can be found in Gudehus (1989); Uson, Boughn \& Kuhn (1991); Mackie (1992); V\'{\i}lchez--G\'omez
et al. (1994a).

\subsubsection{Sky Level.} If we want to fix what is the real extension of our intracluster light, we need a correct 
determination of the sky level. Thus, we ought to be sure that we are far enough of the central part of the cluster to
reach the end of the diffuse light profile. We can get this either working with a big field (i.e., making a tessellation 
of different images as in Scheick \& Kuhn 1994) or studying a relatively distant cluster in order to be sure that all
the cluster is inside our CCD.

\subsubsection{Faint Galaxies.} We have to correct from the contamination due to the galaxies fainter than the 
completeness limit in magnitude for our sample. One possibility is to extrapolate a Schechter luminosity function fitted
to our data, until the detection limit. But if we use the k-correction, then, we have to assume a Hubble type for
this galaxies. If we consider that they are E/S0 galaxies we tend to overcorrect the diffuse light in the red filters
with respect to the blue ones (V\'{\i}lchez--G\'omez et al. 1994a).

\subsubsection{Other Sources of Errors.} For example, vignetting in the image, incorrect determination of the galactic
absorption, statistical errors associate with the measure, wrong redshift for the cluster, the presence of galactic
cirrus as reported by Haikala \& Mattila (1995) or Szomoru \& Guhathakurta (1998).

\section{Characteristics}

I will try to summarize some of the most important characteristics associated with the diffuse light in clusters of
galaxies:

\subsubsection{Luminosity.} It shows a wide range. The intracluster light can represent between the 10\% and the 50\%
of the total light of the region where it is detected. Schombert (1988) finds some correlation, but faint, 
between the luminosity
of the cD envelope and that of the underlying galaxy. This correlation can hint that the process of formation of the brightest
cluster galaxy (BCG) has some reflection in the origin of its envelope. 

\subsubsection{Color.} Different authors have report various results. 
Valentijn (1983) in $B - V$ and Scheick \& Kuhn (1994) in $V - R$ find blueward gradients that vary between 0.1 to 0.6
mag drop. 
Schombert (1988) in $B - V$ doesn't find
any evidence of strong color gradients or blue envelopes colors. Finally, Mackie (1992) in $g - r$ reports a reddening at the 
end of the envelopes, in one case of the order of 0.15 mag.

\subsubsection{Structure.} Schombert (1988) and Mackie, Visvanathan \& Carter (1990) find a apparent break in the 
surface brightness profile of the underlying cD galaxies. According to Schombert (1988), this break is found near the
$24 V$ mag arcsec$^{-2}$ but there are no sharp changes in either eccentricity or orientation between the galaxy and the
envelope. However, Uson et al. (1991) and Scheick \& Kuhn (1994) don't see such a break in their studies.  

Reinforcing the idea of common evolutive processes, Schombert (1988) and Bernstein et al. (1995) find that the 
diffuse light, globular cluster density and galaxy density profiles seem to have similar radial structure, proportional
to $r^{-2.6}$. However, Cl 1613+31 shows a different profile for the diffuse light and the galaxies 
(V\'{\i}lchez--G\'omez et al. 1994a).

\subsubsection{Cluster properties and diffuse light.} Schombert (1988), in one of the most comprehensive studies of 
cD envelopes, finds the following correlations between the luminosity of the envelope ($L_{\rm env}$) 
and the general properties
of the cluster: (1) There is a clear correlation between $L_{\rm env}$ and cluster richness for compact, regular clusters;
(2) there is no evident correlation with velocity dispersion; (3) there is a slight correlation with the Bautz--Morgan or
Rood--Sastry cluster type; (4) there is an unambiguous correlation with the X-ray luminosity.

Finally, there are no reports of galaxies with envelopes in the field and the cD-like galaxies 
observed in poor clusters dwell in local density maxima, comparable to the central region of rich clusters
(West \& Van den Bergh 1991). That is, a cluster or subcluster environment with high local density contrast looks like
an unambiguous
requirement for the presence of cD envelopes or intracluster light.

\section{Sources for the diffuse light}

Basically, there are five processes to explain the origin of the intracluster light:

\subsubsection{Stars from the outer envelopes of galaxies.} Sometimes the extension of the diffuse light is so
large (several core radius) that is hard to believe that these stars are gravitationally bound to any
galaxy, and probably, they are stripped material after the interaction between galaxies. This could be
the case in Cl 1613+31 (V\'{\i}lchez--G\'omez et al. 1994a). Also, it could be that the stars have born
directly in the intergalactic medium from a cooling flow, for example (Prestwich \& Joy 1991).

\subsubsection{Dwarf galaxies and globular clusters.} Part of the light in the intergalactic medium in
distant clusters, where it is not possible to resolve dwarf galaxies and globular clusters, can have this
origin. Nevertheless, Bernstein et al. (1995) have measure in the Coma cluster a diffuse light apart from dwarf galaxies and
globular clusters.

\subsubsection{Hot intracluster bremsstrahlung radiation in the optical.} Woolf (1967), Mattila (1977) and
Bernstein et al. (1995) established that, at least for the Coma cluster, this contribution is not 
significant if we take into account the boundaries imposed by the observations in X-ray and the observed
H$\beta$ intensity.

\subsubsection{Light scattered by intergalactic dust.} The existence of dust in rich clusters of galaxies
as established by Zwicky (1959) or Hu (1992) would suggest the production of diffuse scattered light. 
Mattila (1977) makes an estimation of around 12\% of the total surface brightness of the Coma cluster 
can be due to the surface brightness of the scattered light with origin in the dust.

\subsubsection{The radiative decay of particles.} Partridge (1990) considers that the radiative decay of
low mass particle ($m_{\nu} \sim 4$ eV) would produce extragalactic light in the visible.

\bigskip

The first source seems to be the most important. Scheick \& Kuhn (1994), studying the diffuse light ``granularity''
in Abell 2670 established that the luminosity of each source is less that $10^4$ L$_{\odot}$. This suggests that the 
main origin of the diffuse light is light from stars since the luminosity associated with the sources 
is about a factor 100 smaller than the luminosity
of the faintest dwarf galaxies. Similar result is found by Bernstein et al. (1995) for the Coma cluster.

\section{Origins of the cD envelopes}

There are basically four theories that try to elucidate what is the origin and evolution of the cD envelopes. 
None of them offers a complete picture of the problem.

\subsubsection{Stripping theory.}
 
This theory was initially proposed  by Gallagher \& Ostriker (1972). According with this theory, the origin of the
envelope is on the debris due to tidal interactions between the cluster galaxies. These stars and gas are then deposited
in the potential well of the cluster where the BCG is located. This process begins after the
cluster collapse and the envelope grows as the cluster evolves. 
The fact that different cD envelopes show different color gradients can be explained as the result of different 
tidal interaction histories: in some clusters the tidal interactions involve mainly spirals, but in others, early type
galaxies are the source material (Schombert 1988).
The main problem to this hypothesis is the
difficulty to explain the observed smoothness of the envelopes as the timescale to 
dissolve the clumps is on the order of 
the crossing time of the cluster (Scheick \& Kuhn 1994).

\subsubsection{Primordial origin theory.}

This hypothesis, suggested by Merrit (1984), is similar to the previous one but, in this case, the process of removing
stars from the halos of the galaxies was carried out by the mean cluster tidal field and took place during the initial
collapse of the cluster. The BCG, due to its privileged position in relation with the potential well,
gets the residuals that make up its envelope. However, this picture is difficult to reconciliate with the fact that there
are cD's with significant peculiar velocities (Gebhardt \& Beers 1991) as well as with the 
smoothness of the diffuse light 
either the envelope is affixed to the cD or fixed and the cD is moving through it. Moreover, if the origin of the 
diffuse light is primordial, how can we explain the observation of blue color gradients in some envelopes,
supposed little activity after virialization?

\subsubsection{Cooling flows.}

Fabian \& Nulsen (1977) proposed that the radiative cooling of hot X-ray gas can produce an increase of the densities
around the BCG until star formation takes place. But this process is confined to the first one hundred
kpc from the center of the cluster (Prestwich \& Joy 1991), 
insufficient to explain the big envelopes of several hundred of kpc observed. Moreover,
a blue gradient color is expected if recent star formation is taking place.

\subsubsection{Mergers.}

Villumsen (1982, 1983) found that after a merger with the BCG, and under special conditions, it is possible to form
an halo similar to that present in cD galaxies since there is a transfer of energy to the outer part of the merger,
resulting an extended envelope. Although this theory reproduces the profile observed for the envelopes, it is not
possible to accomplish for the luminosities and masses seen for the diffuse light. However, in poor clusters where 
there are cD-like galaxies without a clear envelope this mechanism can play a more important role (Thuan \& 
Romanishin 1981; Schombert 1986).

\section{Conclusions}

After this review, it is clear that it is necessary to carry out a more systematic study of the diffuse light in clusters
to obtain a better comprehension of the origin and evolution of its properties and its relation with the global 
characteristics of the cluster.

\acknowledgments

I would like to thank R. Pell\'o and B. Sanahuja for their help and comments in my study of the diffuse light in clusters.
I am grateful to Kalevi Mattila for enabling my participation in this conference. I thank also STScI and IAU for 
financial support.


\begin{references}
\reference Abell, G. O. 1965, \araa, 3, 1
\reference Arp, H., \& Bertola, F. 1969, Astrophys. Letters, 4, 23
\reference Arp, H., \& Bertola, F. 1971, \apj, 163, 195
\reference Baum, W. A. 1955, \pasp, 67, 328
\reference Baum, W. A. 1973, \pasp, 85, 530
\reference Bernstein, G. M., Nichol, R. C., Tyson, J. A., Ulmer, M. P., \& Wittman, D. 1995, \aj, 110, 1507
\reference Boughn, S. P., \& Uson, J. M. 1997, \apj, 488, 44
\reference Fabian, A. C., \& Nulsen, P. E. J. 1977, \mnras, 180, 479
\reference Gallagher, J. S., \& Ostriker, J. P. 1972, \aj, 77, 288
\reference Gebhardt, K., \& Beers, T. C. 1991, \apj, 383, 72
\reference Gudehus, D. H. 1989, \apj, 340, 661
\reference Gunn, J. E. 1969, \baas, 1, 191
\reference Gunn, J. E., \& Melnick, J. 1975, \baas, 7, 412
\reference Haikala, L. K., \& Mattila, K. 1995, \apjl, 443, L33
\reference Holmberg, E. 1958, Medd. Luns. Astron. Obs. II, No. 136, 63
\reference Hu, E. M. 1992, \apj, 391, 608
\reference Kormendy, J., \& Bahcall, J. N. 1974, \aj, 79, 671
\reference Mackie, G. 1992, \apj, 400, 65 
\reference Mackie, G., Visvanathan, N., \& Carter, D. 1990, \apj, 73, 637
\reference Matthews, T. A., Morgan, W. W., \& Schmidt, M. 1964, \apj, 140, 35
\reference Mattila, K. 1977, \aap, 60, 425
\reference Melnick, J., White, S. D. M., \& Hoessel, J. 1977, \mnras, 180, 207
\reference Merrit, D. 1984, \apj, 276, 26
\reference Oemler, A. 1973, \apj, 180, 11
\reference Oemler, A. 1976, \apj, 209, 693
\reference Partridge, R. B. 1990, in The Galactic and Extragalactic Background Radiation, S. Bowyer \& C. Leinert,
           Dordrecht: Kluwer, 283
\reference Prestwich, A. H., \& Joy, M. 1991, \apjl, 369, L1
\reference Scheick, X., \& Kuhn, J. R. 1994, \apj, 423, 566
\reference Schombert, J. M. 1986, \apjs, 60, 603
\reference Schombert, J. M. 1988, \apj, 328, 475
\reference Szomoru, A., \& Guhathakurta, P. 1998, \apjl, 494, L93
\reference Thuan, T. X., \& Kormendy J. 1977, \pasp, 89, 466
\reference Thuan, T. X., \& Romanishin, W. 1981, \apj, 248, 439
\reference Uson, J. M., Boughn, S. P., \& Kuhn, J. R. 1991, \apj, 369, 46
\reference Valentijn, E. A. 1983, \aap, 118, 123
\reference Vaucouleurs, G. de 1960, \apj, 131, 585
\reference Vaucouleurs, G. de 1969, Astrophys. Letters, 4, 17
\reference Vaucouleurs, G. de, \& Vaucouleurs, A. de 1970, Astrophys. Letters, 5, 219
\reference V\'{\i}lchez--G\'omez, R., Pell\'o, R., \& Sanahuja, B. 1994a, \aap, 283, 37
\reference V\'{\i}lchez--G\'omez, R., Pell\'o, R., \& Sanahuja, B. 1994b, \aap, 289, 661
\reference Villumsen, J. V. 1982, \mnras, 199, 493
\reference Villumsen, J. V. 1983, \mnras, 204, 219
\reference Welch, G. A., \& Sastry, G. N. 1971, \apjl, 169, L3
\reference Welch, G. A., \& Sastry, G. N. 1972, \apjl, 171, L81
\reference West, M. J., \& Van den Bergh, S. 1991, \apj, 373, 1
\reference Woolf, N. J. 1967, \apj, 148, 287
\reference Zwicky, F. 1951, \pasp, 63, 61 
\reference Zwicky, F. 1957, Morphological Astronomy, Berlin: Springer-Verlag, 48
\reference Zwicky, F. 1959, in Encyclopedia of Physics, S. Fl\"ugge, Berlin: Springer-Verlag, 53, 398
\end{references}
\end{document}